\documentclass[aps,preprint]{revtex4}
\usepackage[T1]{fontenc}
\usepackage[latin1]{inputenc}
\usepackage{amsmath}
\usepackage{graphics}
\usepackage{graphicx}
\usepackage{amssymb}
\usepackage{dcolumn}
\usepackage{bm}
\usepackage{amsfonts}
\begin{document}
\preprint{ }
\title{Search for excited spin-3/2 and spin-1/2 leptons at linear colliders}
\author{O. Cakir }
\email{ocakir@science.ankara.edu.tr}
\author{A. Ozansoy}
\email{aozansoy@science.ankara.edu.tr}
\affiliation{Ankara University, Faculty of Sciences, Department of Physics, 06100,
Tandogan, Ankara,Turkey}

\begin{abstract}
We study the single production of excited spin-3/2 and spin-1/2 leptons in
future high energy $e^{+}e^{-}$ collisions. We calculate the production cross
section and decay widths of excited spin-3/2 and spin-1/2 leptons according to
their effective currents. We show that these possible new excited states can
be probed up to the mass $m^{\ast}\sim\sqrt{s}$ depending on their couplings
to leptons and gauge bosons. We present the angular distributions of final
state particles as a measure to discriminate between an excited spin-3/2 and
spin-1/2 lepton signal. The signals and the corresponding backgrounds are
studied in detail to obtain attainable limits on masses and couplings of
excited leptons at future linear colliders.

\end{abstract}
\date[]{}
\maketitle

\section{Introduction}

The replication of three fundamental fermion generations suggests the
possibility that they are composite structures made up of more fundamental
constituents. \ At a scale of constituent binding energies there should appear
new interactions among quarks and leptons. Such interactions can appear by a
constituent interchange or by exchange of a messenger particle. Excited
leptons and quarks ($l^{\ast},q^{\ast}$) appear as a consequence of the
compositeness \cite{Terazawa77}. In the framework of composite models of
quarks and leptons an excited spin-1/2 lepton is considered to be the lowest
lying radial and orbital excitation, then excited spin-3/2 states are also
expected to exist \cite{Lopes80}. Composite quarks and leptons in the enlarged
groups of the standard theory would also imply spin-3/2 quarks and leptons
\cite{Tosa85}. Further motivation for spin-3/2 particles comes from the
supergravity (SUGRA) where spin-3/2 gravitino is as the superpartner of the
graviton \cite{Freedman76}.

Phenomenologically an excited lepton can be considered to be a heavy lepton
sharing leptonic quantum number with the corresponding ordinary lepton.
Massive spin-3/2 excited states with the analogous heavy spin-1/2 excited one
can be produced at \ future high energy colliders through their effective
interactions with the ordinary leptons. Previous studies of production \ and
decay properties of charged spin-3/2 leptons can be found in \cite{Lopes80,
Choudhury85, Almeida96}. In electron-positron collisions excited spin-3/2
leptons can be produced in pairs, but it is limited kinematically by
\ $m^{\ast}<\sqrt{s}/2$, while single production can reach masses as high as
$\sqrt{s}$. An analysis of the production and decay processes of single heavy
spin-3/2 leptons was performed in \cite{Almeida96} in the frame of two
phenomenological currents taking into account only the signal 
(without background consideration) coupled to $Z$-boson and $W$-boson.

Current direct limits on the masses of excited charged spin-1/2 leptons are:
$m^{\ast}>103.2$ GeV from pair production at LEP experiments \cite{Abbiendi02}
assuming non chiral transition couplings. $m^{\ast}>255$ GeV from single
production at HERA assuming chiral couplings for excited electron $e^{\ast}$
\cite{Adlof02}. Relatively small mass limits for single production of excited
muon $m^{\ast}>190$ GeV \cite{Abbiendi02}, and $m^{\ast}>185$ GeV from single
production of excited tau. The mass limits for excited neutrino from single
production $m^{\ast}>190$ GeV, and $m^{\ast}>102.6$ GeV from pair production
\cite{Achard03} assuming $f=-f^{^{\prime}}$. Relatively small mass limits are
obtained for $f=f^{^{\prime}}$, where $f$ and $f^{^{\prime}}$ are the scaling
factors \ for the gauge couplings of $SU(2)$ and $U(1).$ Indirect limits on
the mass of excited spin-1/2 electron is $m^{\ast}>310$ GeV from LEP
experiment assuming chiral couplings \cite{Achard02, Yao06}.\

In this study, we start with a transition magnetic type couplings between
ordinary and excited spin-1/2 leptons, and three types of phenomenological
currents of the spin-3/2 fields motivated by the new physics scale available
at future high energy collisions. In the upcoming sections we present the
branching rates of excited spin-3/2 and spin-1/2 electrons, the production
cross sections depending on their masses at the available center of mass
energies of future high energy $e^{+}e^{^{\_}}$ colliders, namely
International Linear Collider (ILC) \cite{Loew03} with $\sqrt{s}=0.5$ TeV and
Compact Linear Collider (CLIC) \cite{Assmann00} with an optimal design energy
of $\sqrt{s}=3$ TeV. Finally, we compare the angular distributions of final
state particles to discriminate between spin-3/2 and spin-1/2 excited electron
signal from the corresponding background.

\section{Phenomenological Currents}

\begin{figure}[ptbh]
{{}}\includegraphics[
width=9cm,
height=6cm
]{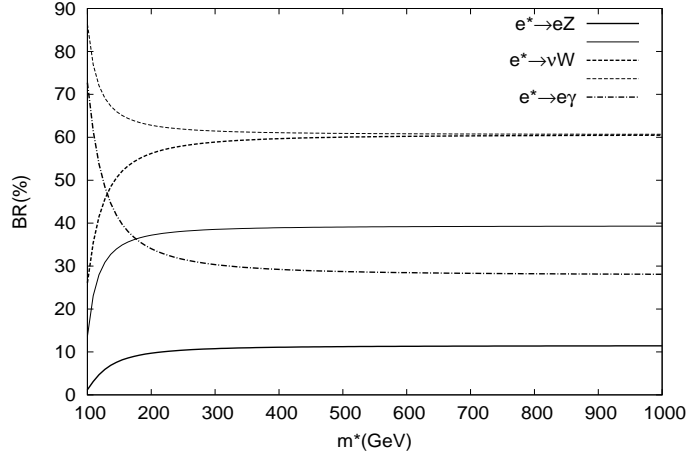}\caption{The branching ratios ($\%$) depending on the mass of
excited spin-1/2 electron for $f=f^{\prime}=1$ (thick lines) for
$f=-f^{\prime}=1$ (thin lines).}
\label{fig1}
\end{figure}

The interaction between a spin-1/2 excited electron, gauge boson
($V=\gamma,Z,W^{\pm}$) and the SM lepton is described by the effective current:%

\begin{equation}
J_{1/2}^{\mu}=\frac{g_{e}}{2\Lambda}\overline{u}(k,1/2)i\sigma^{\mu\nu}q_{\nu
}(1-\gamma_{5})f_{V}u(p,1/2) \label{1}%
\end{equation}
where $\Lambda$ is the scale of new physics responsible for the new
interactions, $k,p$ and $q$ are the four-momentum of SM lepton, excited
spin-1/2 electron and gauge boson, respectively. Electromagnetic coupling
constant is given by $g_{e}=\sqrt{4\pi\alpha}$ .\ $f_{V}$ is the electroweak
coupling parameter corresponding to a vector boson. In Eq. (1) $\sigma^{\mu
\nu}=i(\gamma^{\mu}\gamma^{\nu}-\gamma^{\nu}\gamma^{\mu})/2$ where
$\gamma^{\mu}$ being the Dirac matrices. For an excited spin-1/2 electron,
three decay channels are possible: radiative decay $e^{\ast}\rightarrow
e\gamma$, neutral weak decay $e^{\ast}\rightarrow eZ$, charged weak decay
$e^{\ast}\rightarrow\nu W.$ If we neglect the SM\ lepton masses we find decay
widths as%

\begin{equation}
\Gamma(e^{\ast(1/2)}\longrightarrow lV)=\frac{\alpha m^{\ast3}}{4\Lambda^{2}%
}f_{V}^{2}(1-\frac{m_{V}^{2}}{m^{\ast2}})^{2}(1+\frac{m_{V}^{2}}{2m^{\ast2}})
\end{equation}
where $f_{\gamma}=-(f+f^{\prime})/2$, $f_{Z}=(-f\cot\theta_{W}+f^{\prime}%
\tan\theta_{W})/2$ and $f_{W}=f/\sqrt{2}\sin\theta_{W}$ for excited charged
lepton. The parameters $f$ and $f^{\prime}$ are determined by the composite
dynamics, and they can be changed to $q^{2}-$ dependent form factors. In the
literature, they are often taken as $f=f^{\prime}=1$ or $f=-f^{\prime}=1$ for
$\Lambda=m^{\ast}$. The total decay width of the excited spin-1/2 electron is
$\Gamma\simeq1.9(6.9)$ GeV for $m^{\ast}=0.3(1)$ TeV at $f=f^{\prime}=1$ and
$\Lambda=m^{\ast}$. When we take $f=-f^{\prime}=1$ the results for decay
widths slightly change for smaller mass values, and we find the total decay
width for $e^{\ast}$ as $\Gamma\simeq1.8(6.9)$ GeV at $\Lambda=m^{\ast
}=0.3(1)$ TeV. For these cases the total decay width of $e^{\ast}$ depending
on its mass can be found in Table \ref{table1}. The branching ratios of the
excited spin-1/2 electrons into SM leptons and gauge bosons are given in Fig.
\ref{fig1}. As seen from Fig. \ref{fig1} charged current decays become
dominant for higher mass values $m^{\ast}>150$ GeV. Their relative importance
depends on the gauge boson mass and couplings.

\begin{figure}[ptbh]
{{}}
\par
\begin{center}
\includegraphics[
height=6cm,
width=10cm
]{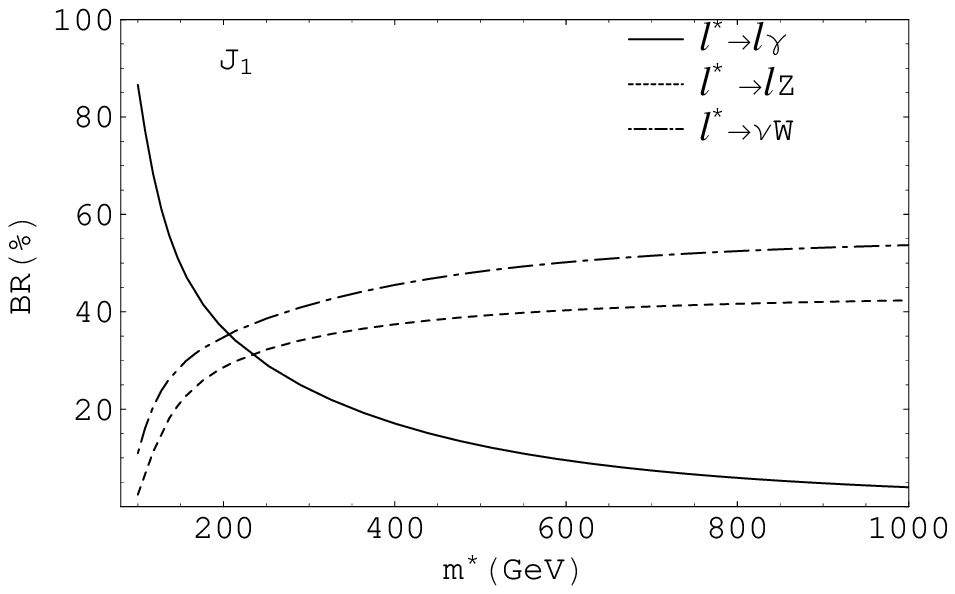}
\end{center}
\par
{}
\par
\begin{center}
\includegraphics[
height=6cm,
width=10cm
]{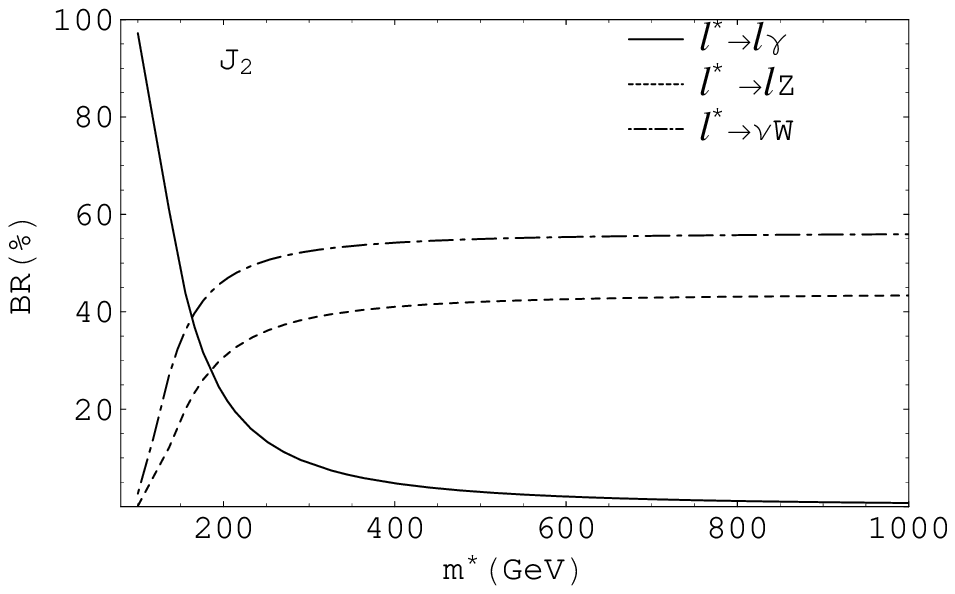}
\end{center}
\par
{}
\par
\begin{center}
\includegraphics[
height=6cm,
width=10cm
]{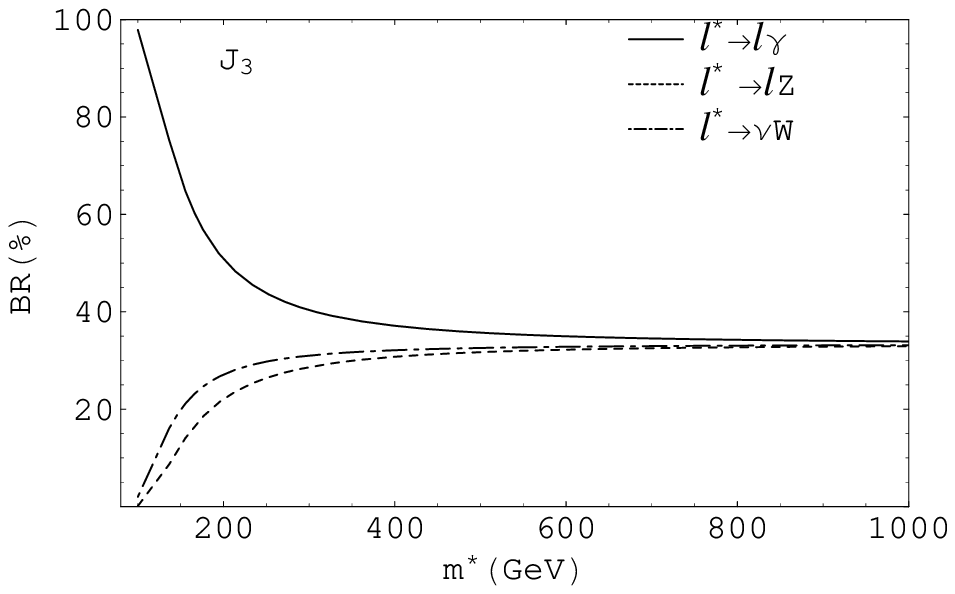}
\end{center}
\caption{The branching ratios as functions of excited spin-3/2 electron mass.
The plots are given separately for $J_{1}$, $J_{2}$ and $J_{3}$ currents.}
\label{fig2}
\end{figure}

The phenomenological currents for the interactions among spin-3/2 excited
electron, gauge boson and SM lepton are given by

\begin{equation}
J_{1}^{\mu}=g_{e}\bar{u}(k,1/2)(c_{1V}-c_{1A}\gamma_{5})u^{\mu}(p,3/2)
\end{equation}

\begin{equation}
J_{2}^{\mu}=\frac{g_{e}}{\Lambda}\bar{u}(k,1/2)q_{\lambda}\gamma^{\mu}%
(c_{2V}-c_{2A}\gamma_{5})u^{\lambda}(p,3/2)
\end{equation}

\begin{equation}
J_{3}^{\mu}=\frac{g_{e}}{\Lambda^{2}}\bar{u}(k,1/2)q_{\lambda}i\sigma^{\mu\nu
}q_{\nu}(c_{3V}-c_{3A}\gamma_{5})u^{\lambda}(p,3/2)
\end{equation}
where $u^{\mu}(p,3/2)$ represents the Rarita-Schwinger vector-spinor
\cite{Schwinger41}, $c_{iV}$ and $c_{iA}$ are the vector and axial vector
couplings, the four momenta $q_{\alpha}=(p-k)_{\alpha}$. A spin-3/2 excited
electron ($e^{\ast}$) decays via two-body process according to the
phenomenological currents. The weak decay widths of excited spin-3/2 electrons
for three different currents are given by

\begin{align}
\Gamma_{1}(e^{\ast(3/2)}  &  \longrightarrow lV)=\frac{\alpha}{48}(c_{1V}
^{2}+c_{1A}^{2})m^{\ast}\frac{(1-\kappa)^{2}}{\kappa}(1+10\kappa+\kappa^{2})\\
\Gamma_{2}(e^{\ast(3/2)}  &  \longrightarrow lV)=\frac{\alpha}{48}(c_{2V}
^{2}+c_{2A}^{2})m^{\ast}(\frac{m^{\ast}}{\Lambda})^{2}\frac{(1-\kappa)^{4}
}{\kappa}(1+2\kappa)\\
\Gamma_{3}(e^{\ast(3/2)}  &  \longrightarrow lV)=\frac{\alpha}{48}(c_{3V}
^{2}+c_{3A}^{2})m^{\ast}(\frac{m^{\ast}}{\Lambda})^{4}(1-\kappa)^{4}(2+\kappa)
\end{align}
where $\kappa=(m_{V}/m^{\ast})^{2}$ and $m_{V\text{ }}$ is the mass of the
vector ($W^{\pm}$ or $Z$) boson. The radiative decay widths of excited
spin-3/2 electrons are given as

\begin{align}
\Gamma_{1}(e^{\ast(3/2)}  &  \longrightarrow e\gamma)=\frac{\alpha}{4}
(c_{1V}^{\gamma^{2}}+c_{1A}^{\gamma^{2}})m^{\ast}\\
\Gamma_{2}(e^{\ast(3/2)}  &  \longrightarrow e\gamma)=\frac{\alpha}{24}
(c_{2V}^{\gamma^{2}}+c_{2A}^{\gamma^{2}})m^{\ast}(\frac{m^{\ast}}{\Lambda
})^{2}\\
\Gamma_{3}(e^{\ast(3/2)}  &  \longrightarrow e\gamma)=\frac{\alpha}{48}
(c_{3V}^{\gamma^{2}}+c_{3A}^{\gamma^{2}})m^{\ast}(\frac{m^{\ast}}{\Lambda
})^{4}%
\end{align}
Here, one may note that the dimension five and six operators contributes as
$\Lambda^{-2}$ and $\Lambda^{-4}$ to the decay widths, their relative
importance is not essential when $\Lambda=m^{\ast}$. Taking $\Lambda=m^{\ast
}=0.5(1)$ TeV and $c_{iV}=c_{iA}=0.5$ we find the total decay widths of the
excited spin-3/2 electrons as $\Gamma_{1}=3.9(24.6)$ GeV, $\Gamma
_{2}=2.7(22.2)$ GeV and $\Gamma_{3}=0.23(0.48)$ GeV for the currents $J_{1}$,
$J_{2}$ and $J_{3}$, respectively. The decay widths of spin-3/2 $e^{\ast}$ are
shown in Table \ref{table1}. At equal couplings and $\Lambda=m^{\ast}$, the
difference between the decay widths is due to the $\kappa$ terms and different
radiative contributions for each currents. The corresponding branching ratios
are given in Fig. \ref{fig2}. For equal couplings and $\Lambda=m^{\ast}$ the
branching ratios for the weak decays corresponding to the current $J_{1}$ and
$J_{2}$ appear to be dominant for $m^{\ast}\gtrsim200$ GeV. For the current
$J_{3}$ radiative and weak decay channels with the same couplings have equal
probability for large $m^{\ast}$.

\begin{table}[tbp] \centering
\caption{Decay widths of excited spin-3/2 electrons with $c_V=c_A=0.5$ depending on their mass values.
Numbers in parenthesis show for the case $f=-f'=1$ for spin-1/2 excited electron.
\label{table1}}{{}}
\begin{tabular}
[c]{l|l|l|l|l|}\hline
$m^{\ast}$(TeV) & $\Gamma_{J(1/2)}$ (GeV) & $\Gamma_{J_{1}(3/2)}$ (GeV) &
$\Gamma_{J_{2}(3/2)}$ (GeV) & $\Gamma_{J_{3}(3/2)}$ (GeV)\\\hline
$0.2$ & $1.15$ $(1.03)$ & $0.54$ & $0.14$ & $0.06$\\\hline
$0.3$ & $1.93$ $(1.85)$ & $1.22$ & $0.55$ & $0.12$\\\hline
$0.4$ & $2.67$ $(2.61)$ & $2.29$ & $1.36$ & $0.18$\\\hline
$0.5$ & $3.39$ $(3.35)$ & $3.89$ & $2.71$ & $0.23$\\\hline
$0.75$ & $5.18$ $(5.15)$ & $11.12$ & $9.31$ & $0.36$\\\hline
$1.0$ & $6.95$ $(6.93)$ & $24.62$ & $22.20$ & $0.48$\\\hline
$1.5$ & $10.47(10.45)$ & $78.89$ & $75.24$ & $0.73$\\\hline
$2.0$ & $13.98(13.97)$ & $183.48$ & $178.61$ & $0.97$\\\hline
$2.5$ & $17.49(17.47)$ & $355.16$ & $349.07$ & $1.22$\\\hline
$3.0$ & $20.99(20.98)$ & $650.72$ & $603.41$ & $1.46$\\\hline
\end{tabular}
\end{table}

\section{Cross Sections}

\begin{figure}[ptbh]
{{}}\includegraphics[
height=2cm,
width=7cm
]{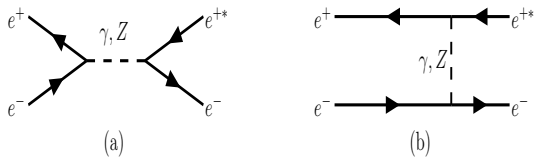}\caption{Feynman diagrams for excited spin-3/2 or spin-1/2 positron
production in $e^{+}e^{-}$ collisions. }
\label{fig3}
\end{figure}

The high energy $e^{+}e^{-}$ collisions provide an excellent environment to
search for excited leptons. Spin-3/2 and spin-1/2 excited leptons can be
produced at future $e^{+}e^{-}$ colliders, namely ILC and CLIC. The Feynman
diagrams for single production of excited positron (or electron) via the
$s$-channel and $t$-channel $\gamma,Z$ exchange are shown in Fig. \ref{fig3}.
On the other hand, single and pair production of excited muon or tau in the
$s$-channel are also possible while their production in the t-channel leads to flavor changing neutral currents (FCNC).

We calculate the cross sections for $e^{+}e^{-}\rightarrow
e^{+\ast}e^{-}$ process for three currents $J_{1},J_{2}$ and $J_{3}$.
Recently, spin-1/2 excited electron and neutrino have been studied at the ILC
energy $\sqrt{s}=0.5$ TeV in \cite{Cakir04}. The $e^{\ast}\rightarrow e\gamma$
and $\nu^{\ast}\rightarrow eW$ signal have been searched for an easy
identification and accurate measurements in a linear collider environment. The
results show that spin-1/2 excited electron (neutrino) can be probed up to the
mass $m^{\ast}=350(450)$ GeV in the radiative (charged weak) decay channel
when the coupling parameters are taken as $f=f^{\prime}=0.1$ for
$\Lambda=m^{\ast}$.

\begin{figure}[ptbh]
{{}}\includegraphics[
height=6cm,
width=10cm
]{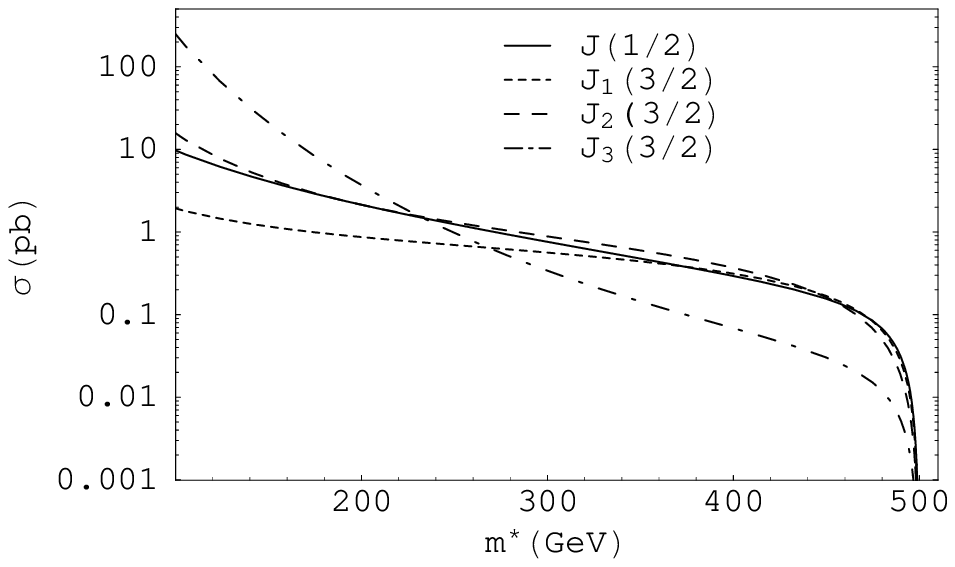} 
\includegraphics[
height=6cm,
width=10cm]{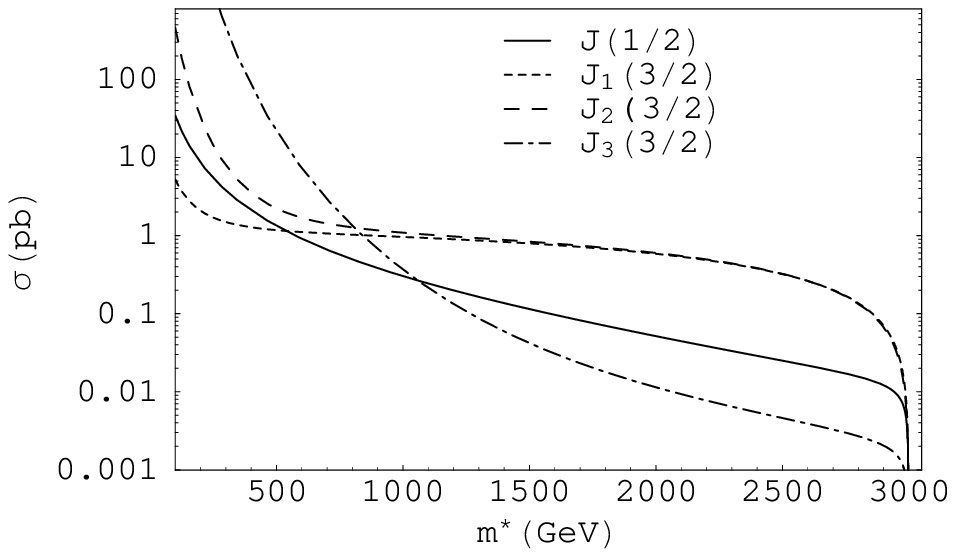}
\caption{Excited electron production cross section as a function
of the mass at ILC energy $\sqrt{s}=0.5$ TeV (upper) and CLIC\ energy
$\sqrt{s}=3$ TeV (lower). Solid, dotted, dashed and dot-dashed lines denote
spin-1/2 and spin-3/2 with $J_{1},$ $J_{2}$, $J_{3}$ currents for
$f=-f^{\prime}=1$ and $c_{iv}^{Z}=c_{iA}^{Z}=0.5$, respectively.}
\label{fig4}
\end{figure}

The explicit calculations of the diagrams for excited spin-3/2 electrons lead to

\begin{equation}
\frac{d\sigma}{dt}=\frac{{{g}}_{{e}}^{2\,}}{24{{m}}^{\ast2}\,\pi\,s^{2}
}\underset{i,j=1,4}{\sum}\frac{T_{ij}}{P_{ij}}
\end{equation}
where we use the expressions $T_{ij}$ and $P_{ij}$ as given in the Appendix.
Total cross sections as a function of excited electron mass are shown in Fig.
\ref{fig4} at $\sqrt{s}=0.5$ and $3$ TeV.

One can see from Fig. \ref{fig4} excited spin-3/2 electrons with current
$J_{3}$ has a cross section larger than the other two currents when its mass
below $1$ TeV when produced at $\sqrt{s}=3$ TeV. Depending on the couplings
$c_{iV},c_{iA}$ the relative importance of three currents become more
pronounced compared to the excited spin-1/2 electrons. In the analysis, we
take into account only one coupling (i.e. coupling to one of the gauge bosons) that
is kept free while the others are assumed to vanish. In fact, there is no
theoretical prediction on the total cross section for spin-3/2 single
production. The $Z$-boson exchange in the $s$- and $t$-channel acts a heavy
propagator as an effective form factor. On the other hand, if we choose
$f=-f^{\prime}$ photon decouples from the excited spin-1/2 electrons.

\begin{figure}[ptbh]
{{}}\includegraphics[
height=6cm,
width=10cm
]{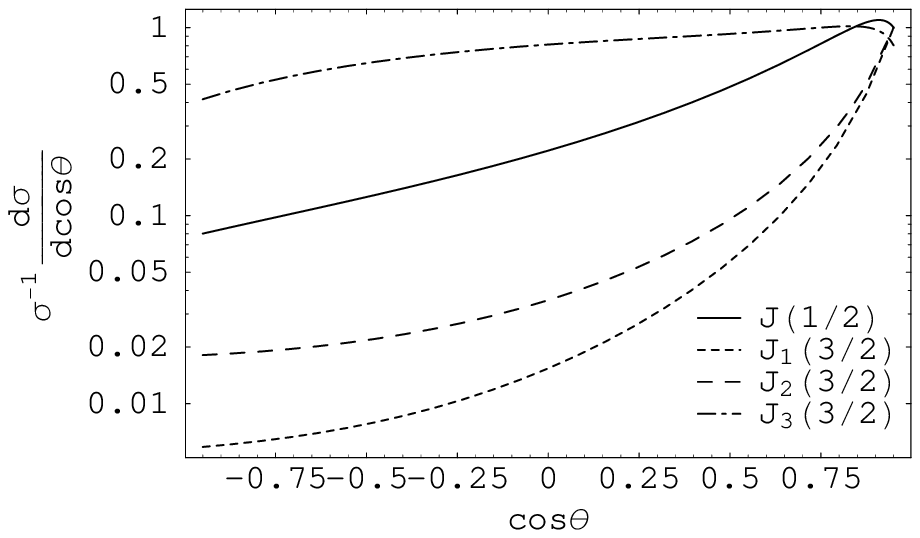} \includegraphics[
height=6cm,
width=10cm
]{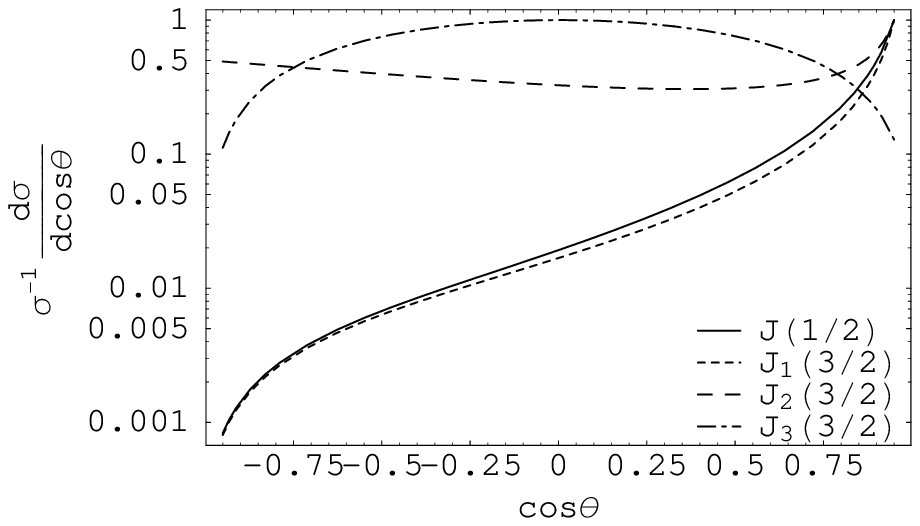}\caption{The differential cross section as a function of the
scattering angle for different spin-3/2 currents, and spin-1/2 excited
electrons at (upper) $\sqrt{s}=0.5$ TeV and (lower) $\sqrt{s}=3$ TeV.}
\label{fig5}
\end{figure}

In order to differentiate the spin-3/2 and spin-1/2 excited electron signals
we plot normalized differential cross sections as a function of $\cos\theta$
in Fig. \ref{fig5}. The excited spin-1/2 charged lepton is produced mostly in
the forward direction, while excited spin-3/2 lepton shows different angular
shape for the current $J_{3}$ than the others. For equal couplings, i.e.
$c_{iV}=c_{iA},$ the normalized cross sections are not coupling dependent. For
the final state we consider three decay channels of signal $e^{\ast
}\rightarrow e\gamma$, $e^{\ast}\rightarrow eZ$ and $e^{\ast}\rightarrow\nu W$
for excited electron. In the first and second decay channels we obtain clear
signal for a discovery, while the last decay channel for $e^{\ast}$ has
uncertainty due to the neutrino in the final state.

In order to perceive the excited electron signals from the background we put
kinematical cuts on the final state particles. The signal can be more
pronounced over the background by applying suitable cuts. We consider the
processes $e^{+}e^{-}\rightarrow e^{-}e^{+}\gamma$ , $e^{+}e^{-}\rightarrow
e^{-}e^{+}Z$ and $e^{+}e^{-}\rightarrow e^{-}\overline{\nu}W^{+}$ for the
background. We calculate the background cross sections by using CalcHEP
\cite{Pukhov99}. We apply the following acceptance cuts: $p_{T}^{e,\gamma}>20$
GeV, $\left\vert \eta_{e,\gamma}\right\vert <2.5$, $\Delta R_{(e^{+}%
e^{-}),(e^{\pm}\gamma)}>0.4$, where $p_{T}$ is the transverse momentum of
final state detectable particle, $\eta$ denotes pseudo rapidity and $\Delta R$
is the separation between two of them. The corresponding backgrounds are
studied by applying the cuts on the transverse momentum and their
pseudo rapidities of final state leptons and jets or missing transverse
momentum when the SM neutrino is produced. After applying these cuts we find
the cross section for the SM background as $1.93(0.16)$ pb, $0.11(0.03)$ pb
and $0.92(0.46)$ pb at $\sqrt{s}=0.5(3)$TeV for $e^{-}e^{+}\gamma$,
$e^{-}e^{+}Z$ and $e^{-}\overline{\nu}W^{+}$, respectively. Furthermore, a way
of extracting the excited electron signal from the background is to impose the
cut $\left\vert m_{e\gamma}-m^{\ast}\right\vert <25$ GeV on the invariant mass
of the $e\gamma$ system and $\left\vert m_{eZ}-m^{\ast}\right\vert <25$ GeV on
the $eZ$ invariant mass for the neutral weak decay channels of $e^{\ast}$.
These cuts can be relaxed for higher values of the masses and couplings of
$e^{\ast}$. For $m^{\ast}>1.5$ TeV we apply a cut $m_{eV}>1$ TeV for an easy
detection of the signal. We present signal and background cross sections in
Table \ref{table2} and \ref{table3} for different mass values taking equal couplings.

\begin{table}[tbp] \centering
\caption{The cross sections for the signal and background after the cuts
at $e^+e^-$ colliders with 0.5 TeV. The signal cross section is given for
$\Lambda=m^*$ and $c^Z_{iV}=c^Z_{iA}=0.5$ for spin-3/2 and $f=-f'=1$
for spin-1/2 excited electron. \label{table2}}{{}}
\begin{tabular}
[c]{l|llll|l}\hline
&  &  & $\sigma_{S}$(pb) &  & \\\hline
$m^{\ast}$(GeV) & $J(1/2)$ & \multicolumn{1}{|l}{$J_{1}(3/2)$} &
\multicolumn{1}{|l}{$J_{2}(3/2)$} & \multicolumn{1}{|l|}{$J_{3}(3/2)$} &
$\sigma_{B}$(pb)\\\hline
$200$ & $2.14$ & \multicolumn{1}{|l}{$0.87$} & \multicolumn{1}{|l}{$2.13$} &
\multicolumn{1}{|l|}{$3.68$} & $1.47\times10^{-2}$\\\hline
$250$ & $1.24$ & \multicolumn{1}{|l}{$0.69$} & \multicolumn{1}{|l}{$1.31$} &
\multicolumn{1}{|l|}{$0.97$} & $1.27\times10^{-2}$\\\hline
$300$ & $0.76$ & \multicolumn{1}{|l}{$0.56$} & \multicolumn{1}{|l}{$0.88$} &
\multicolumn{1}{|l|}{$0.34$} & $1.26\times10^{-2}$\\\hline
$350$ & $0.48$ & \multicolumn{1}{|l}{$0.44$} & \multicolumn{1}{|l}{$0.60$} &
\multicolumn{1}{|l|}{$0.14$} & $1.44\times10^{-2}$\\\hline
$400$ & $0.29$ & \multicolumn{1}{|l}{$0.31$} & \multicolumn{1}{|l}{$0.37$} &
\multicolumn{1}{|l|}{$0.07$} & $1.73\times10^{-2}$\\\hline
$475$ & $0.08$ & \multicolumn{1}{|l}{$0.08$} & \multicolumn{1}{|l}{$0.07$} &
\multicolumn{1}{|l|}{$0.02$} & $9.73\times10^{-3}$\\\hline
\end{tabular}
\end{table}

\begin{table}[tbp] \centering
\caption{The cross sections for the signal and background after the cuts
at $e^+e^-$ colliders with 3 TeV. The signal cross section is given for
$\Lambda=m^*$ and $c^Z_{iV}=c^Z_{iA}=0.5$ for spin-3/2 and $f=-f'=1$
for spin-1/2 excited electron. \label{table3}}{{}}
\begin{tabular}
[c]{l|llll|l}\hline
&  &  & $\sigma_{S}$(pb) &  & \\\hline
$m^{\ast}$(GeV) & $J(1/2)$ & \multicolumn{1}{|l}{$J_{1}(3/2)$} &
\multicolumn{1}{|l}{$J_{2}(3/2)$} & \multicolumn{1}{|l|}{$J_{3}(3/2)$} &
$\sigma_{B}$(pb)\\\hline
$250$ & $5.45$ & \multicolumn{1}{|l}{$1.69$} & \multicolumn{1}{|l}{$14.4$} &
\multicolumn{1}{|l|}{$1384.2$} & $1.67\times10^{-4}$\\\hline
$500$ & $1.33$ & \multicolumn{1}{|l}{$1.17$} & \multicolumn{1}{|l}{$2.22$} &
\multicolumn{1}{|l|}{$21.71$} & $8.26\times10^{-4}$\\\hline
$1000$ & $0.30$ & \multicolumn{1}{|l}{$0.96$} & \multicolumn{1}{|l}{$1.08$} &
\multicolumn{1}{|l|}{$0.37$} & $3.24\times10^{-4}$\\\hline
$1500$ & $0.11$ & \multicolumn{1}{|l}{$0.79$} & \multicolumn{1}{|l}{$0.84$} &
\multicolumn{1}{|l|}{$0.04$} & $3.67\times10^{-4}$\\\hline
$2000$ & $0.05$ & \multicolumn{1}{|l}{$0.06$} & \multicolumn{1}{|l}{$0.06$} &
\multicolumn{1}{|l|}{$0.01$} & $1.66\times10^{-2}$\\\hline
$2500$ & $0.025$ & \multicolumn{1}{|l}{$0.032$} & \multicolumn{1}{|l}{$0.032$}
& \multicolumn{1}{|l|}{$0.004$} & $1.66\times10^{-2}$\\\hline
$2750$ & $0.017$ & \multicolumn{1}{|l}{$0.17$} & \multicolumn{1}{|l}{$0.17$} &
\multicolumn{1}{|l|}{$0.003$} & $1.66\times10^{-2}$\\\hline
\end{tabular}
\end{table}

\begin{figure}[ptbh]
{{}}\includegraphics[
height=6cm,
width=10cm
]{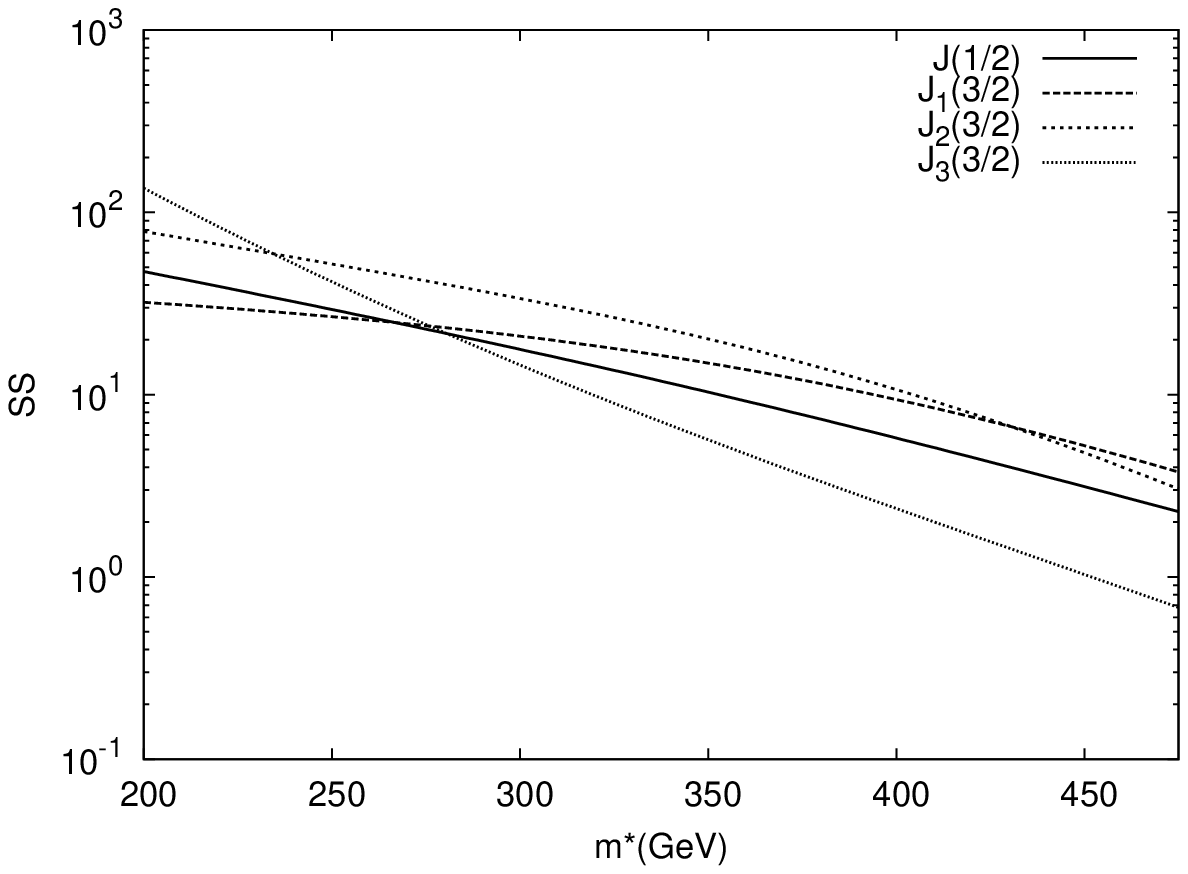} \includegraphics[
height=6cm,
width=10cm
]{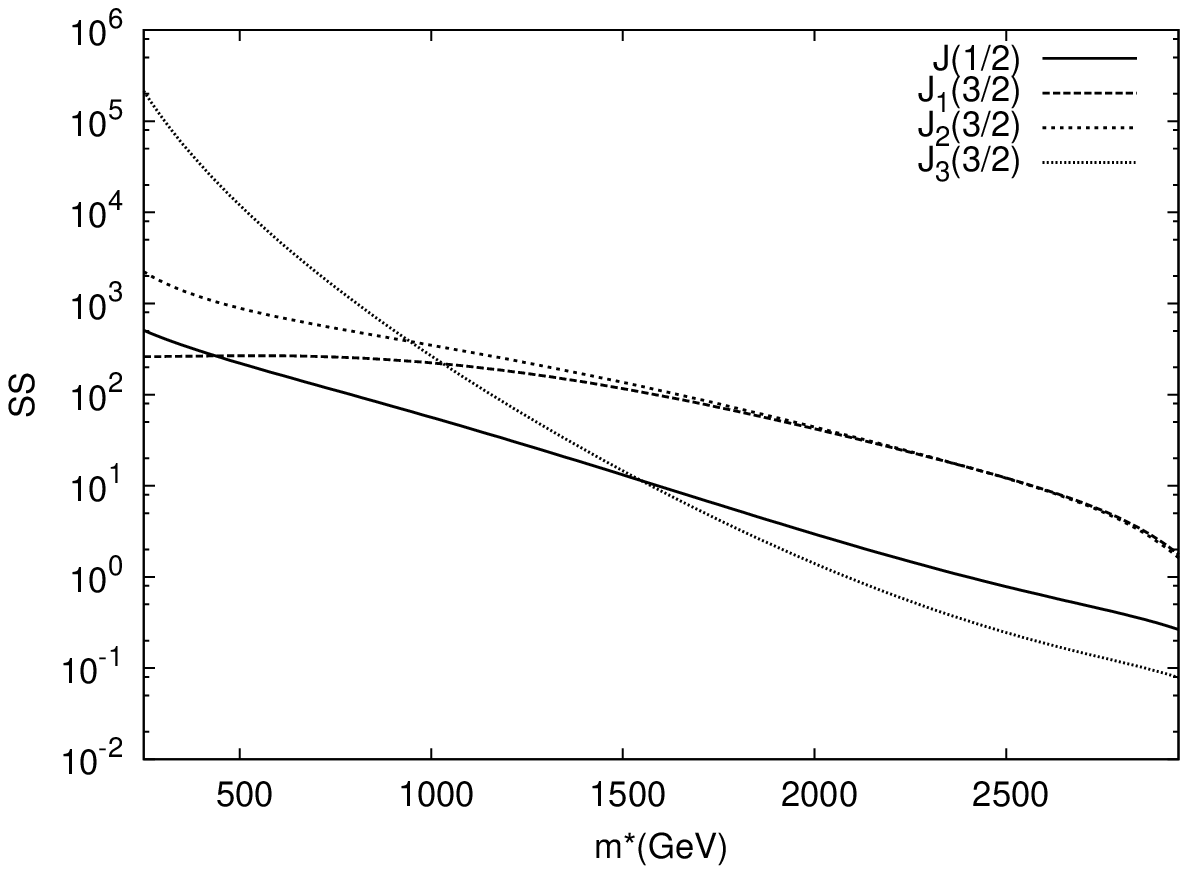}\caption{Statistical significance for the excited electron signal
(both for spin-1/2 and spin-3/2) at linear $e^{+}e^{-}$ colliders with (upper)
$\sqrt{s}=0.5$ TeV and (lower) $\sqrt{s}=3$ TeV. }
\label{fig6}
\end{figure}

\begin{figure}[ptbh]
{{}}\includegraphics[
height=6cm,
width=10cm
]{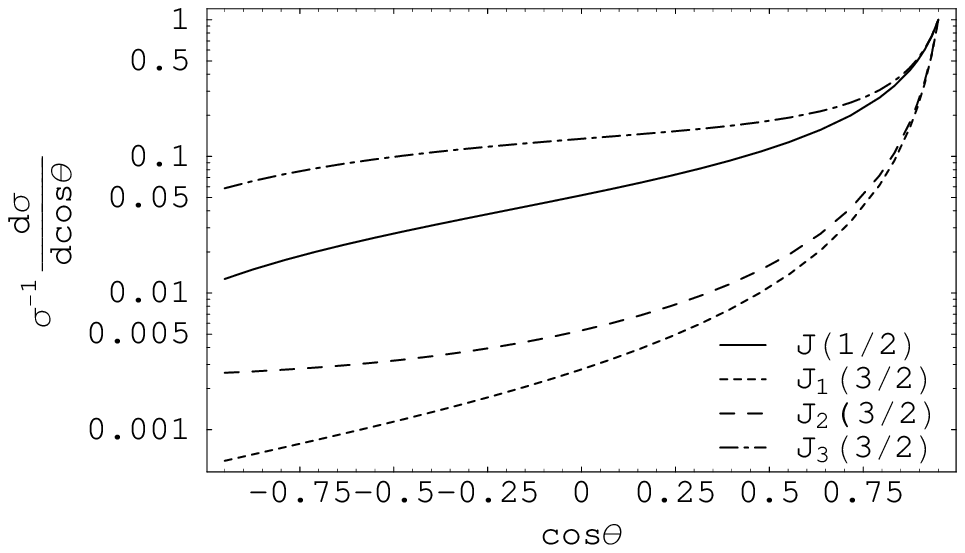} \includegraphics[
height=6cm,
width=10cm
]{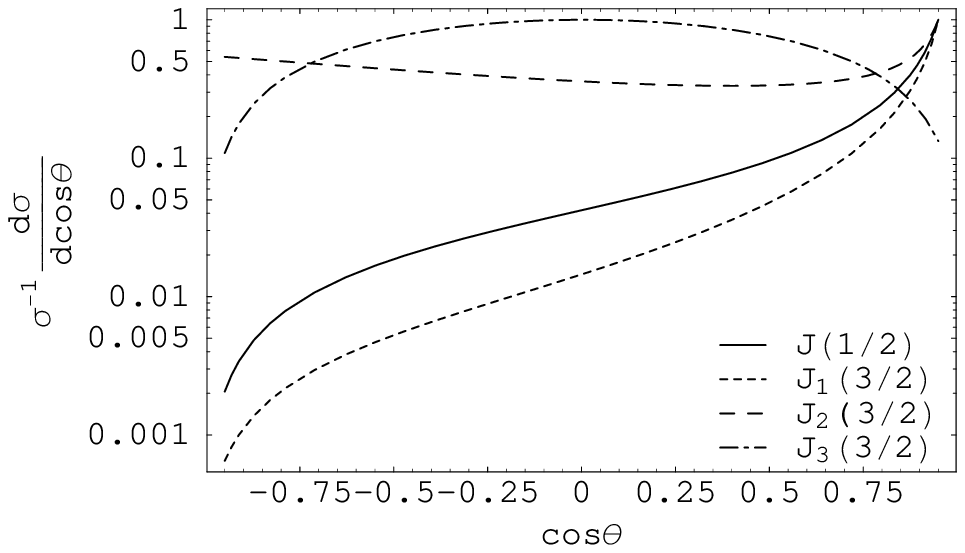}
\caption{The differential cross section as a function of the
scattering angle for different spin-3/2 currents, and spin-1/2 excited
electrons at (upper) $\sqrt{s}=0.5$ TeV and (lower) $\sqrt{s}=3$ TeV.}
\label{fig7}
\end{figure}

\begin{figure}[ptbh]
{{}}\includegraphics[
height=6cm,
width=10cm
]{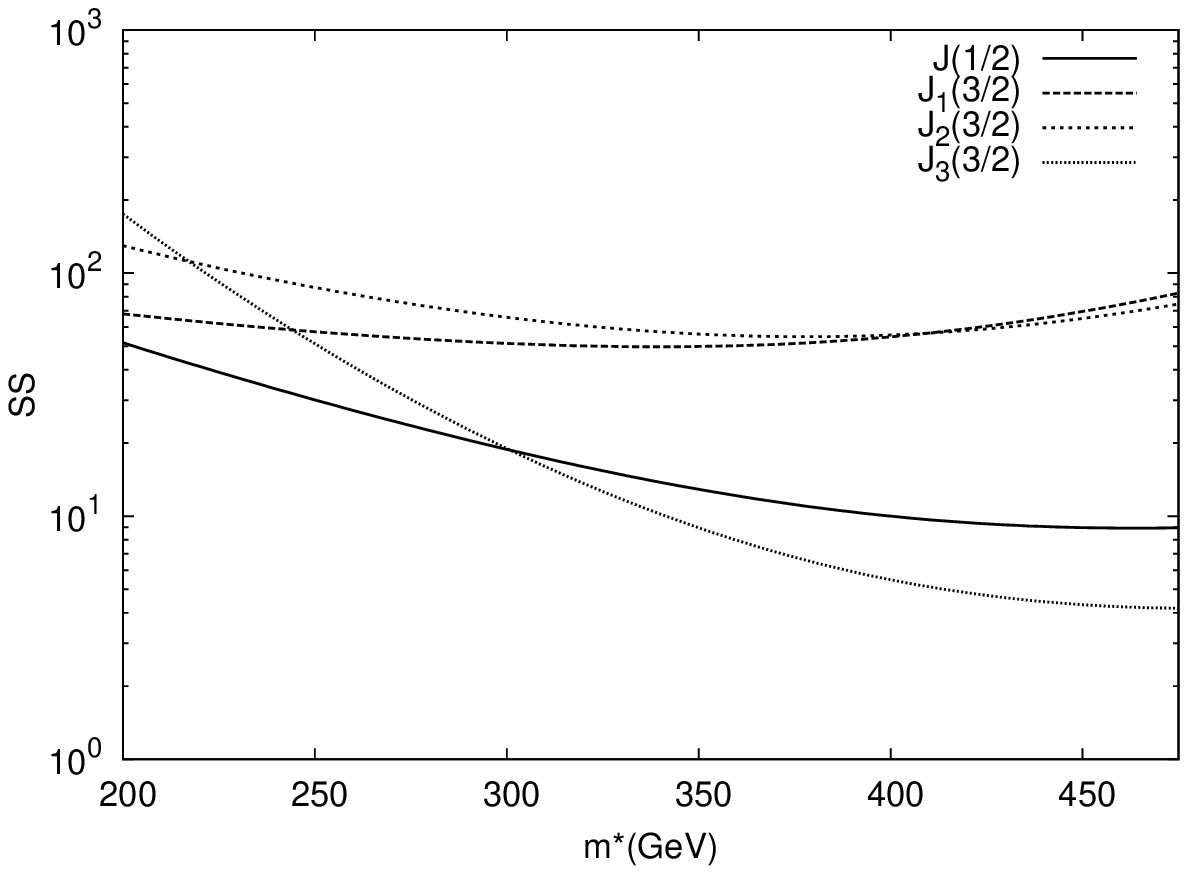} \includegraphics[
height=6cm,
width=10cm
]{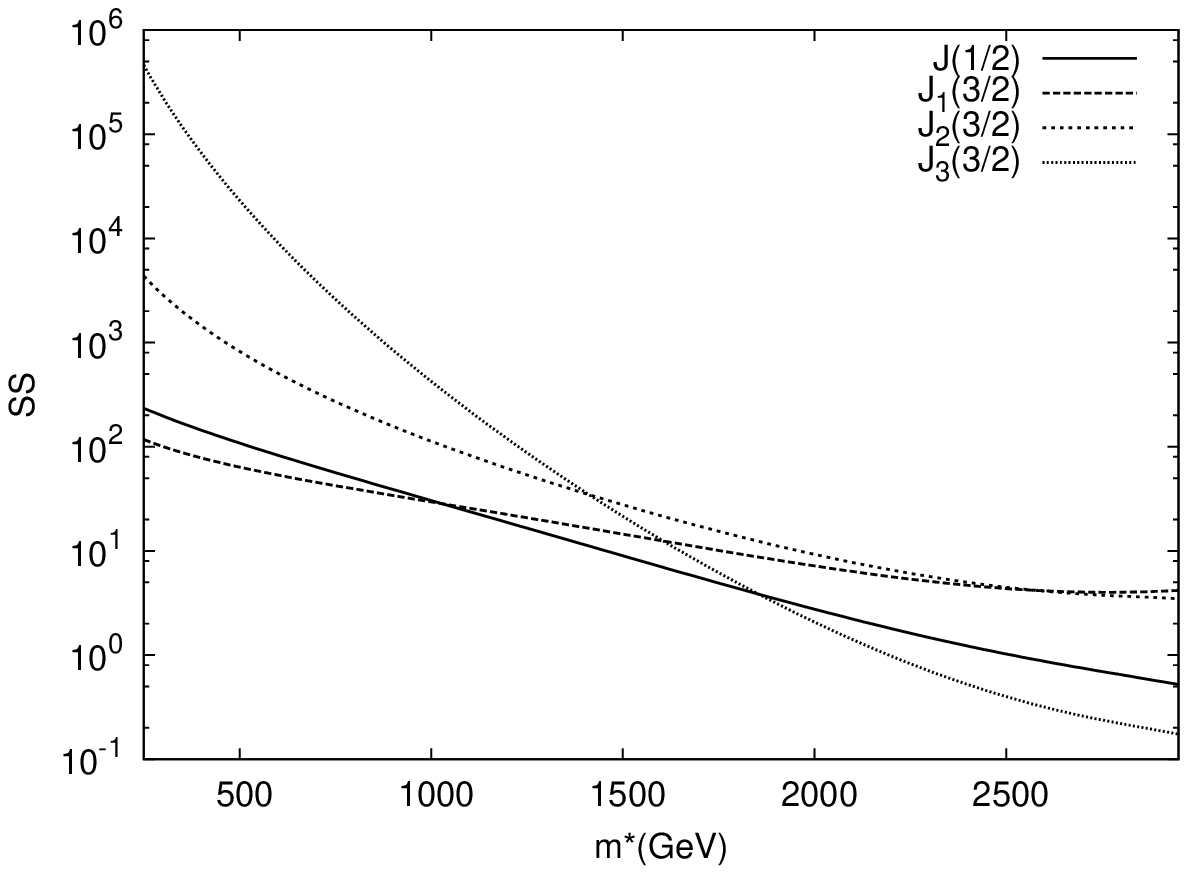}\caption{The differential cross section as a function of the
scattering angle for different spin-3/2 currents, and spin-1/2 excited
electrons at (upper) $\sqrt{s}=0.5$ TeV and (lower) $\sqrt{s}=3$ TeV.}%
\label{fig8}
\end{figure}

Fig. \ref{fig6} shows the cross section for excited spin-3/2 and spin-1/2 electrons, when they have electromagnetic couplings, depending on its mass for the ILC and CLIC energies. In order to determine the couplings $c_{iV}^\gamma,c_{iA}^\gamma$
we study the signal $e^*\to e\gamma$ and the corresponding background
by applying the above mentioned cuts.
In Table \ref{table4} and \ref{table5}, we present the cross sections
for signal and background for the chosen mass bin intervals
at $\sqrt{s}=0.5$ TeV and $3$ TeV. Excited positrons (or electrons) with spin-3/2 can be distinguished from the spin-1/2 signal by searching for the angular distribution of associated electrons (or positrons) as shown in Fig. \ref{fig7}. Concerning the criteria $SS\geq3$, and taking the couplings $c_V=c_A=0.05$ the excited spin-3/2 electrons (having current $J_3$) can be observed up to 0.48 TeV and 2.0 TeV at ILC and CLIC energies
as shown in Fig. \ref{fig8}. As can be seen from Fig. \ref{fig8}, for the other currents accessible range for the masses becomes enlarged.

\begin{table}[tbp] \centering
\caption{The cross sections for the signal and background after the cuts
at $e^+e^-$ colliders with 0.5 TeV. The signal cross section is given for
$\Lambda=m^*$ and $c^\gamma_{iV}=c^\gamma_{iA}=0.5$ for spin-3/2 and $f=f'=1$
for spin-1/2 excited electron. \label{table4}}{{}}
\begin{tabular}
[c]{l|llll|l}\hline
&  &  & $\sigma_{S}$(pb) &  & \\\hline
$m^{\ast}$(GeV) & $J(1/2)$ & \multicolumn{1}{|l}{$J_{1}(3/2)$} &
\multicolumn{1}{|l}{$J_{2}(3/2)$} & \multicolumn{1}{|l|}{$J_{3}(3/2)$} &
$\sigma_{B}$(pb)\\\hline
$200$ & $3.61$ & \multicolumn{1}{|l}{$4.75$} & \multicolumn{1}{|l}{$9.06$} &
\multicolumn{1}{|l|}{$12.28$} & $9.76\times10^{-2}$\\\hline
$250$ & $2.22$ & \multicolumn{1}{|l}{$4.27$} & \multicolumn{1}{|l}{$6.28$} &
\multicolumn{1}{|l|}{$3.46$} & $1.17\times10^{-1}$\\\hline
$300$ & $1.48$ & \multicolumn{1}{|l}{$4.07$} & \multicolumn{1}{|l}{$5.05$} &
\multicolumn{1}{|l|}{$1.33$} & $1.37\times10^{-1}$\\\hline
$350$ & $1.03$ & \multicolumn{1}{|l}{$4.00$} & \multicolumn{1}{|l}{$4.41$} &
\multicolumn{1}{|l|}{$0.65$} & $1.52\times10^{-1}$\\\hline
$400$ & $0.74$ & \multicolumn{1}{|l}{$4.02$} & \multicolumn{1}{|l}{$4.05$} &
\multicolumn{1}{|l|}{$0.37$} & $1.49\times10^{-1}$\\\hline
$475$ & $0.45$ & \multicolumn{1}{|l}{$4.16$} & \multicolumn{1}{|l}{$3.75$} &
\multicolumn{1}{|l|}{$0.21$} & $5.05\times10^{-2}$\\\hline
\end{tabular}
\end{table}

\begin{table}[tbp] \centering
\caption{The cross sections for the signal and background after the cuts
at $e^+e^-$ colliders with 3 TeV. The signal cross section is given for
$\Lambda=m^*$ and $c^\gamma_{iV}=c^\gamma_{iA}=0.5$ for spin-3/2 and $f=-f'=1$
for spin-1/2 excited electron. \label{table5}}{{}}
\begin{tabular}
[c]{l|llll|l}\hline
&  &  & $\sigma_{S}$(pb) &  & \\\hline
$m^{\ast}$(GeV) & $J(1/2)$ & \multicolumn{1}{|l}{$J_{1}(3/2)$} &
\multicolumn{1}{|l}{$J_{2}(3/2)$} & \multicolumn{1}{|l|}{$J_{3}(3/2)$} &
$\sigma_{B}$(pb)\\\hline
$250$ & $2.52$ & \multicolumn{1}{|l}{$1.04$} & \multicolumn{1}{|l}{$38.54$} &
\multicolumn{1}{|l|}{$4095.9$} & $3.15\times10^{-3}$\\\hline
$500$ & $0.63$ & \multicolumn{1}{|l}{$0.33$} & \multicolumn{1}{|l}{$2.93$} &
\multicolumn{1}{|l|}{$63.5$} & $1.75\times10^{-3}$\\\hline
$1000$ & $0.15$ & \multicolumn{1}{|l}{$0.15$} & \multicolumn{1}{|l}{$0.37$} &
\multicolumn{1}{|l|}{$0.99$} & $1.37\times10^{-3}$\\\hline
$1500$ & $0.07$ & \multicolumn{1}{|l}{$0.12$} & \multicolumn{1}{|l}{$0.17$} &
\multicolumn{1}{|l|}{$0.10$} & $9.19\times10^{-4}$\\\hline
$2000$ & $0.04$ & \multicolumn{1}{|l}{$0.11$} & \multicolumn{1}{|l}{$0.13$} &
\multicolumn{1}{|l|}{$0.02$} & $3.31\times10^{-2}$\\\hline
$2500$ & $0.03$ & \multicolumn{1}{|l}{$0.11$} & \multicolumn{1}{|l}{$0.11$} &
\multicolumn{1}{|l|}{$0.009$} & $3.31\times10^{-2}$\\\hline
$2750$ & $0.02$ & \multicolumn{1}{|l}{$0.11$} & \multicolumn{1}{|l}{$0.11$} &
\multicolumn{1}{|l|}{$0.007$} & $3.31\times10^{-2}$\\\hline
\end{tabular}
\end{table}

For the analysis of the excited spin-3/2 and spin-1/2 electrons at linear colliders we
define the statistical significance ($SS$) of the signal as

\[
SS=\frac{\sigma_{S}}{\sqrt{\sigma_{B}}}\sqrt{\epsilon.L}
\]
where $L_{int}$ is the integrated luminosity of the collider and $\epsilon$ is
the efficiency to detect the signal in the chosen channel. In Fig.
\ref{fig6}, discovery reach for spin-3/2 excited electrons are shown. With an
integrated luminosity of $2\times10^{5}$ pb$^{-1}$ the ILC\ can probe excited
spin-3/2 electrons up to the mass $400$ GeV if only the current $J_{3}$ is
realized for $c_{3V}^{Z}=c_{3A}^{Z}=0.05$. For the same couplings other
currents gives higher observability for the signal. At CLIC with $\sqrt{s}=3$
TeV excited electrons can be probed up to the mass $2.0$ TeV if their
couplings are taken as $c_{3V}^{Z}=c_{3A}^{Z}=0.05$.

\section{Conclusion}

We consider only the gauge interactions of excited spin-3/2 and spin-1/2
electrons with the SM particles. In principle, excited leptons can also
couples to SM leptons via contact interactions which may enlarge the discovery
limits for the future colliders \cite{Baur90,Cakir03}. \ If excited fermions
having spin-1/2 or spin-3/2 are discovered singly or pairs at the forthcoming
high energy colliders, there will be good reasons to separate them to
understand the underlying dynamics which they obey. Our analysis show that
spin-3/2 excited leptons can be easily separated from the \ spin-1/2 ones by
examining the normalized angular distributions in their single productions.

If polarized $e^{+}$ and $e^{-\text{ }}$beams are used the chiral structure of
the couplings can be identified and more precise measurements can be performed
at the ILC and CLIC. Furthermore, if excited leptons are produced at resonance
in $e\gamma$ collisions, which is an option for the linear colliders, the
angular distribution of the decay products in a frame that they are at rest
give valuable information about their spins. The prospects for the measurement
of the spin of excited leptons will be shown elsewhere \cite{Cakir07}.

Excited leptons can come in three family, $e^{\ast},$ $\mu^{\ast}$ and
$\tau^{\ast},$ here we studied on the excited electron. The study for
$e^{\ast}$ can be enlarged by applying similar analysis to the excited muons
and taus, and their neutrinos at the future high energy linear colliders.

\section{Appendix}

In equation (12) the T-terms $T_{ij}$ and P-terms $P_{ij}$ for the spin-3/2
excited electron interaction current $J_{1}$ are given as

\begin{center}
\begin{align*}
T_{11}^{(1)}  &  =-{{g}}_{{e}}^{2\,}({c}_{1A}^{\gamma}{}^{2}+{c}_{1V}^{\gamma
}{}^{2})({{m}}^{\ast2}-s)\,\left(  -t\,\left(  s+t\right)  +{{m}}^{\ast
2}\,\left(  2\,s+t\right)  \right) \\
T_{12}^{(1)}  &  =[-{{g}}_{{e}}^{\,}\,{{g}}_{{{z}}}(m_{z}^{2}-s)({c}%
_{1A}^{\gamma}({c}_{1V}^{Z}\,{c}_{A}^{f}{{m}}^{\ast2}({{m}}^{\ast2}-s-2t)\\
&  +{c}_{1A}^{Z}{c}_{V}^{f}(st(s+t)+{{m}}^{\ast4}(2s+t)-{{m}}^{\ast2}%
(2s^{2}+2st+t^{2}))+\\
&  {c}_{1V}^{\gamma}({c}_{1A}^{Z}\,{c}_{A}^{f}({{m}}^{\ast2}-s-2t)s+{c}%
_{1V}^{Z}{c}_{V}^{f}(st(s+t)+{{m}}^{\ast4}(2s+t)\\
&  -{{m}}^{\ast2}(2s^{2}+2st+t^{2})))]/2\\
T_{13}^{(1)}  &  =[-{{g}}_{{e}}^{2\,}({c}_{1A}^{\gamma}{}^{2}+{c}_{1V}%
^{\gamma}{}^{2})({{m}}^{\ast4}+st)(s+t)-{{m}}^{\ast2}(s^{2}+t^{2})]/2\\
T_{14}^{(1)}  &  =[{{g}}_{{e}}^{\,}\,{{g}}_{{{z}}}({c}_{A}^{f}({c}%
_{1V}^{\gamma}{c}_{1A}^{Z}+{c}_{1A}^{\gamma}{c}_{1V}^{Z})\\
&  -{c}_{V}^{f}\,({c}_{1A}^{\gamma}{c}_{1A}^{Z}+{c}_{1V}^{\gamma}{c}_{1V}%
^{Z}))(({{m}}^{\ast4}+st)(s+t)-{{m}}^{\ast2}(s^{2}+t^{2}))]/4\\
T_{22}^{(1)}  &  =[-{{g}}_{{{z}}}^{2}{{(4}c}_{1A}^{Z}\,{c}_{1V}^{Z}\,{c}%
_{A}^{f}{c}_{V}^{f}\,{{m}}^{\ast2}s({{m}}^{\ast2}-s-2t)+\\
&  ({c}_{1A}^{Z}{}^{2}+{c}_{1V}^{Z}{}^{2})({c}_{A}^{f2}+{c}_{V}^{f2})\left(
{{m}}^{\ast2}-s\right)  (-t(s+t)+{{m}}^{\ast2}(2s+t)))]/4\\
T_{23}^{(1)}  &  =[{{g}}_{{e}}^{\,}\,{{g}}_{{{z}}}{{(}}\,{c}_{A}^{f}({c}%
_{1V}^{\gamma}{c}_{1A}^{Z}+{c}_{1A}^{\gamma}{c}_{1V}^{Z})-\\
&  {c}_{V}^{f}\,({c}_{1A}^{\gamma}{c}_{1A}^{Z}+{c}_{1V}^{\gamma}{c}_{1V}%
^{Z})\left(  {{m}}_{{{z}}}^{2}-s\right)  (({{m}}^{\ast4}+st)(s+t)-{{m}}%
^{\ast2}(s^{2}+t^{2}))]/4\\
T_{24}^{(1)}  &  =[-\,{{g}}_{{{z}}}^{2}{{(-4}c}_{1A}^{Z}\,{c}_{1V}^{Z}%
\,{c}_{A}^{f}{c}_{V}^{f}\,+\\
&  ({c}_{1A}^{Z}\,^{2}+{c}_{1V}^{Z}\,^{2})({c}_{A}^{f2}+{c}_{V}^{f2})({{m}%
}^{\ast4}(3s-t)+st(s+t)+{{m}}^{\ast2}(-3s^{2}+t^{2})))]/8\\
T_{34}^{(1)}  &  =[-{{g}}_{{e}}^{\,}\,{{g}}_{{{z}}}(({c}_{1A}^{\gamma}%
({c}_{1V}^{Z}\,{c}_{A}^{f}{{m}}^{\ast2}({{m}}^{\ast2}-2s-t)t\\
&  +{c}_{1A}^{Z}{c}_{V}^{f}(-{{m}}^{\ast2}+t)(s(s+t)-{{m}}^{\ast2}(s+2t))))\\
&  +({c}_{1V}^{\gamma}({c}_{1A}^{Z}\,{c}_{A}^{f}{{m}}^{\ast2}({{m}}^{\ast
2}-2s-t)t\\
&  +{c}_{1V}^{Z}{c}_{V}^{f}(-{{m}}^{\ast2}+t)(s(s+t)-{{m}}^{\ast
2}(s+2t)))))]/2\\
T_{33}^{(1)}  &  =-{{g}}_{{e}}^{2\,}({c}_{1A}^{\gamma}{}^{2}+{c}_{1V}^{\gamma
}{}^{2})({{m}}^{\ast2}-t)\,\left(  -s\,\left(  s+t\right)  +{{m}}^{\ast
2}\,\left(  \,s+2t\right)  \right) \\
T_{44}^{(1)}  &  =[{{g}}_{{{z}}}^{2}(4{c}_{1A}^{Z}\,{c}_{1V}^{Z}\,{c}_{A}%
^{f}{c}_{V}^{f}\,{{m}}^{\ast2}\,t\,\left(  -{{m}}^{\ast2}+2s+\,t\right)  -\\
&  ({c}_{1A}^{Z}{}^{2}+{c}_{1V}^{Z}{}^{2})({c}_{A}^{f2}+{c}_{V}^{f2})({{m}%
}^{\ast2}-t)(-s(s+t)+{{m}}^{\ast2}(s+2t))]/4\\
&  \text{and}\\
P_{11}^{(1)}  &  =s^{2}\text{, \ }P_{12}^{(1)}=s(\left(  {{m}}_{{{z}}}%
^{2}-s\right)  ^{2}+{{m}}_{{{z}}}^{2}\Gamma_{z}^{2}),\text{ }P_{13}%
^{(1)}=st,\text{ }P_{14}^{(1)}=s({{m}}_{{{z}}}^{2}-t),\text{ }\\
P_{22}^{(1)}  &  =({{m}}_{{{z}}}^{4}+s^{2}+{{m}}_{{{z}}}^{2}(-2s+\Gamma
_{z}^{2})),\\
P_{23}^{(1)}  &  =t({{m}}_{{{z}}}^{4}+s^{2}+{{m}}_{{{z}}}^{2}(-2s+\Gamma
_{z}^{2})),\text{ }P_{24}^{(1)}=\left(  {{m}}_{{{z}}}^{2}-t\right)  ({{m}%
}_{{{z}}}^{4}+s^{2}+{{m}}_{{{z}}}^{2}(-2s+\Gamma_{z}^{2})),\\
P_{34}^{(1)}  &  =\left(  {{m}}_{{{z}}}^{2}-t\right)  t,\text{ }\\
P_{33}^{(1)}  &  =t^{2},\text{ }P_{44}^{(1)}=\left(  {{m}}_{{{z}}}%
^{2}-t\right)  ^{2}%
\end{align*}

The $T_{ij}$ and $P_{ij}$ terms for the current $J_{2}$ are given by%

\begin{align*}
T_{11}^{(2)}  &  =[-{{g}}_{{e}}^{2\,}({c}_{2A}^{\gamma}{}^{2}+{c}_{2V}%
^{\gamma}{}^{2})({{m}}^{\ast2}-s)^{2}\,\left(  -s^{2}-2st-2t^{2}+{{m}}^{\ast
2}\,\left(  s+2t\right)  \right)  ]/2\Lambda^{2}\\
T_{12}^{(2)}  &  =[{{g}}_{{e}}^{\,}\,{{g}}_{{{z}}}(m^{\ast2}-s)^{2}%
(s-m_{Z}^{2})(({c}_{2A}^{\gamma}({c}_{2V}^{Z}\,{c}_{A}^{f}s({{m}}^{\ast
2}-s-2t)\\
&  +{c}_{2A}^{Z}{c}_{V}^{f}(-s^{2}-2st-2t^{2}+m^{\ast2}(s+2t)))\\
&  +{c}_{2V}^{\gamma}({c}_{2A}^{Z}\,{c}_{A}^{f}s({{m}}^{\ast2}-s-2t)+{c}%
_{2V}^{Z}{c}_{V}^{f}(-s^{2}-2st-2t^{2}+m^{\ast2}(s+2t)))]/4\Lambda^{2}\\
T_{13}^{(2)}  &  =[-{{g}}_{{e}}^{2\,}({c}_{2A}^{\gamma}{}^{2}+{c}_{2V}%
^{\gamma}{}^{2})(m^{\ast2}-s-t)(s+t)(m^{\ast4}-st-m^{\ast2}(s+t))]/2\Lambda
^{2}\\
T_{14}^{(2)}  &  =[-{{g}}_{{e}}^{\,}\,{{g}}_{{{z}}}({c}_{A}^{f}({c}%
_{2V}^{\gamma}{c}_{2A}^{Z}+{c}_{2A}^{\gamma}{c}_{2V}^{Z})\\
&  +{c}_{V}^{f}\,({c}_{2A}^{\gamma}{c}_{2A}^{Z}+{c}_{2V}^{\gamma}{c}_{2V}%
^{Z}))(m^{\ast2}-s-t)(s+t)(m^{\ast4}-st-m^{\ast2}(s+t)]/4\Lambda^{2}\\
T_{22}^{(2)}  &  =[-{{g}}_{{{z}}}^{2}\left(  {{m}}^{\ast2}-s\right)
^{2}{{(-4}c}_{2A}^{Z}\,{c}_{2V}^{Z}\,{c}_{A}^{f}{c}_{V}^{f}\,s(-{{m}}^{\ast
2}+s+2t)+\\
&  ({c}_{2A}^{Z}{}^{2}+{c}_{2V}^{Z}{}^{2})({c}_{A}^{f2}+{c}_{V}^{f2})\left(
-s^{2}-2st-2t^{2}+{{m}}^{\ast2}\,\left(  s+2t\right)  \right)  )]/8\Lambda
^{2}\\
T_{23}^{(2)}  &  =[-{{g}}_{{e}}^{\,}\,{{g}}_{{{z}}}{{(}}\,{c}_{A}^{f}({c}%
_{2V}^{\gamma}{c}_{2A}^{Z}+{c}_{2A}^{\gamma}{c}_{2V}^{Z})+{c}_{V}^{f}%
\,({c}_{2A}^{\gamma}{c}_{2A}^{Z}+{c}_{2V}^{\gamma}{c}_{2V}^{Z})\left(  {{m}%
}_{{{z}}}^{2}-s\right) \\
&  (s+t)(-m^{\ast2}+s+t)(-{{m}}^{\ast4}+st+m^{\ast2}(s+t)))]/4\Lambda^{2}\\
T_{24}^{(2)}  &  =[-\,{{g}}_{{{z}}}^{2}{{(4}c}_{2A}^{Z}\,{c}_{2V}^{Z}\,{c}%
_{A}^{f}{c}_{V}^{f}\,+({c}_{2A}^{Z}\,^{2}+{c}_{2V}^{Z}\,^{2})({c}_{A}^{f2}%
+{c}_{V}^{f2})\left(  {{m}}_{{{z}}}^{2}-s\right) \\
&  (-{{m}}^{\ast2}+s)(-2{{m}}^{\ast4}s+{{m}}^{\ast4}(s-t)+st(s+t)+{{m}}%
^{\ast2}(s+t)^{2}))]/8\Lambda^{2}\\
T_{34}^{(2)}  &  =[{{g}}_{{e}}^{\,}\,{{g}}_{{{z}}}({c}_{2A}^{\gamma}({c}%
_{2V}^{Z}\,{c}_{A}^{f}({{m}}^{\ast2}-2s-t)t+{c}_{2A}^{Z}{c}_{V}^{f}%
(-2s^{2}-2st-t^{2}+m^{\ast2}(2s+t)))\\
&  +({c}_{2V}^{\gamma}({c}_{2A}^{Z}\,{c}_{A}^{f}({{m}}^{\ast2}-2s-t)t+{c}%
_{2V}^{Z}{c}_{V}^{f}(-2s^{2}-2st-t^{2}+m^{\ast2}(2s+t))))]/4\Lambda^{2}\\
T_{33}^{(2)}  &  =[-{{g}}_{{e}}^{2\,}({c}_{2A}^{\gamma}{}^{2}+{c}_{2V}%
^{\gamma}{}^{2})({{m}}^{\ast2}-t)^{2}\,(-2s^{2}-2st-t^{2}+m^{\ast
2}(2s+t))]/2\Lambda^{2}\\
T_{44}^{(2)}  &  =[-{{g}}_{{{z}}}^{2}({{m}}^{\ast2}-t)^{2}(4{c}_{2A}^{Z}%
\,{c}_{2V}^{Z}\,{c}_{A}^{f}{c}_{V}^{f}\,\left(  {{m}}^{\ast2}-2s-\,t\right)
t\\
&  +({c}_{2A}^{Z}{}^{2}+{c}_{2V}^{Z}{}^{2})({c}_{A}^{f2}+{c}_{V}^{f2}%
)(-2s^{2}-2st-t^{2}+m^{\ast2}(2s+t))]/8\Lambda^{2}\\
&  \text{and}\\
P_{11}^{(2)}  &  =s^{2}\text{, \ }P_{12}^{(2)}=(m_{z}^{4}s+s^{3}+{{m}}_{{{z}}%
}^{2}(-2s^{2}+\Gamma_{z}^{2})),\text{ }P_{13}^{(2)}=st,\text{ }\\
P_{14}^{(2)}  &  =s({{m}}_{{{z}}}^{2}-t),\text{ }P_{22}^{(2)}=4({{m}}_{{{z}}%
}^{4}+s^{2}+{{m}}_{{{z}}}^{2}s(-2s+\Gamma_{z}^{2})),\\
P_{23}^{(2)}  &  =t({{m}}_{{{z}}}^{4}+s^{2}+{{m}}_{{{z}}}^{2}(-2s+\Gamma
_{z}^{2})),\text{ }P_{24}^{(2)}=\left(  {{m}}_{{{z}}}^{2}-t\right)  ({{m}%
}_{{{z}}}^{4}+s^{2}+{{m}}_{{{z}}}^{2}(-2s+\Gamma_{z}^{2})),\\
P_{34}^{(2)}  &  =\,\left(  -{{m}}_{{{z}}}^{2}+t\right)  t,\text{ }\\
P_{33}^{(2)}  &  =t^{2},\text{ }P_{44}^{(2)}=\left(  {{m}}_{{{z}}}%
^{2}-t\right)  ^{2}%
\end{align*}

The $T_{ij}$ and $P_{ij}$ terms for the current $J_{3}$ are given by%

\begin{align*}
T_{11}^{(3)}  &  =[-{{g}}_{{e}}^{2\,}({c}_{3A}^{\gamma}{}^{2}+{c}_{3V}%
^{\gamma}{}^{2})(m^{\ast2}-s)^{2}\,(m^{\ast4}+2t(s+t)-m^{\ast2}%
(s+2t))]/2\Lambda^{4}\\
T_{12}^{(3)}  &  =[-{{g}}_{{e}}^{\,}\,{{g}}_{{{z}}}(m^{\ast2}-s)^{2}%
(s-m_{Z}^{2})(({c}_{3A}^{\gamma}({c}_{3V}^{Z}\,{c}_{A}^{f}{{m}}^{\ast2}({{m}%
}^{\ast2}-s-2t)\\
&  -{c}_{3A}^{Z}{c}_{V}^{f}(m^{\ast4}+2t(s+t)-m^{\ast2}(s+2t)))\\
&  +{c}_{3V}^{\gamma}({c}_{3A}^{Z}\,{c}_{A}^{f}{{m}}^{\ast2}s({{m}}^{\ast
2}-s-2t)-{c}_{3V}^{Z}{c}_{V}^{f}(m^{\ast4}+2t(s+t)-m^{\ast2}(s+2t)))]/4\Lambda
^{2}\\
T_{13}^{(3)}  &  =[{{g}}_{{e}}^{2\,}({c}_{3A}^{\gamma}{}^{2}+{c}_{3V}^{\gamma
}{}^{2})(m^{\ast2}-s-t)ts]/2\Lambda^{4}\\
T_{14}^{(3)}  &  =[{{g}}_{{e}}^{\,}\,{{g}}_{{{z}}}({c}_{A}^{f}({c}%
_{3V}^{\gamma}{c}_{3A}^{Z}+{c}_{3A}^{\gamma}{c}_{3V}^{Z})-{c}_{V}^{f}%
\,({c}_{3A}^{\gamma}{c}_{3A}^{Z}+{c}_{3V}^{\gamma}{c}_{3V}^{Z}))(m^{\ast
2}-s-t)t(st)]/4\Lambda^{4}\\
T_{22}^{(3)}  &  =[-{{g}}_{z}^{2}\left(  m^{\ast2}-s\right)  ^{2}{{(4}c}%
_{3A}^{Z}\,{c}_{3V}^{Z}\,{c}_{A}^{f}{c}_{V}^{f}\,{{m}}^{\ast2}s(-{{m}}^{\ast
2}+s+2t)+\\
&  ({c}_{3A}^{Z}{}^{2}+{c}_{3V}^{Z}{}^{2})({c}_{A}^{f2}+{c}_{V}^{f2})\left(
{{m}}^{\ast4}+2s(s+t)-m^{\ast2}(2s+t)\right)  )]/8\Lambda^{4}\\
T_{23}^{(3)}  &  =[{{g}}_{{e}}^{\,}\,{{g}}_{{{z}}}s^{2}{{(}}\,{c}_{A}^{f}%
({c}_{3V}^{\gamma}{c}_{3A}^{Z}+{c}_{3A}^{\gamma}{c}_{3V}^{Z})-{c}_{V}%
^{f}\,({c}_{3A}^{\gamma}{c}_{3A}^{Z}+{c}_{3V}^{\gamma}{c}_{3V}^{Z})\\
&  \left(  {{m}}_{{{z}}}^{2}-s\right)  t(-{{m}}^{\ast2}+s+t)]/4\Lambda^{4}\\
T_{24}^{(3)}  &  =[-\,{{g}}_{{{z}}}^{2}{{(-4}c}_{3A}^{Z}\,{c}_{3V}^{Z}%
\,{c}_{A}^{f}{c}_{V}^{f}\,+({c}_{3A}^{Z}\,^{2}+{c}_{3V}^{Z}\,^{2})({c}%
_{A}^{f2}+{c}_{V}^{f2})\\
&  \left(  {{m}}_{{{z}}}^{2}-s\right)  s^{2}t^{2}(-{{m}}^{\ast2}%
+s+t))]/8\Lambda^{4}\\
T_{34}^{(3)}  &  =[-{{g}}_{{e}}^{\,}\,{{g}}_{{{z}}}(m^{\ast4}-6m^{\ast
2}t+t^{2})({c}_{3A}^{\gamma}({c}_{3V}^{Z}\,{c}_{A}^{f}m^{\ast2}(-m^{\ast
2}+2s+t)\\
&  +{c}_{3A}^{Z}{c}_{V}^{f}(m^{\ast4}+2s(s+t)-m^{\ast2}(2s+t)))\\
&  +({c}_{3V}^{\gamma}({c}_{3A}^{Z}\,{c}_{A}^{f}m^{\ast2}(-m^{\ast2}%
+2s+t)+{c}_{3V}^{Z}{c}_{V}^{f}(m^{\ast4}+2s(s+t)-m^{\ast2}(2s+t))))]/4\Lambda
^{4}\\
T_{33}^{(3)}  &  =[-{{g}}_{{e}}^{2\,}({c}_{3A}^{\gamma}{}^{2}+{c}_{3V}%
^{\gamma}{}^{2})(m^{\ast2}-t)^{2}\,(m^{\ast4}+2s(s+t)-m^{\ast2}%
(2s+t))]/2\Lambda^{4}\\
T_{44}^{(3)}  &  =[-{{g}}_{{{z}}}^{2}(4{c}_{3A}^{Z}\,{c}_{3V}^{Z}\,{c}_{A}%
^{f}{c}_{V}^{f}m^{\ast2}\,(m^{\ast2}-t)^{2}\left(  -m^{\ast2}+2s+\,t\right)
t\\
&  +({c}_{3A}^{Z}{}^{2}+{c}_{3V}^{Z}{}^{2})({c}_{A}^{f2}+{c}_{V}^{f2}%
)(m^{\ast4}+2s(s+t)-m^{\ast2}(2s+t)))]/8\Lambda^{4}\\
&  \text{and}\\
P_{11}^{(3)}  &  =s\text{, \ }P_{12}^{(3)}=(m_{z}^{4}s+s^{3}+{{m}}_{{{z}}}%
^{2}(-2s^{2}+\Gamma_{z}^{2})),\text{ }P_{13}^{(3)}=1,\\
\text{ }P_{14}^{(3)}  &  =(t-{{m}}_{{{z}}}^{2}),\text{ }P_{22}^{(3)}=(({{m}%
}_{{{z}}}^{2}-s)^{2}+{{m}}_{{{z}}}^{2}\Gamma_{z}^{2}),\\
P_{23}^{(3)}  &  =(({{m}}_{{{z}}}^{2}-s)^{2}+{{m}}_{{{z}}}^{2}\Gamma_{z}%
^{2}),\text{ }P_{24}^{(3)}=\left(  {{m}}_{{{z}}}^{2}-t\right)  ({{m}}_{{{z}}%
}^{4}+s^{2}+{{m}}_{{{z}}}^{2}(-2s+\Gamma_{z}^{2})),\\
P_{34}^{(3)}  &  =\left(  {{m}}_{{{z}}}^{2}-t\right)  ,\text{ }\\
P_{33}^{(3)}  &  =t,\text{ }P_{44}^{(3)}=\left(  {{m}}_{{{z}}}^{2}-t\right)
^{2}%
\end{align*}

\end{center}

\begin{acknowledgments}
This work was partially supported by the Turkish State Planning Organization
(DPT) under grant No DPT-2006K-120470 and the Turkish Atomic Energy Authority
(TAEK) under grant No VII-B.04.DPT.1.05.
\end{acknowledgments}

\end{document}